\documentclass[pra,twocolumn,showpacs,amsmath,amssymb,superscriptaddress,nofootinbib]{revtex4}

\usepackage{graphicx}

\begin{document}
\title{Quantum teleportation scheme by selecting one of multiple output ports}
\author{Satoshi Ishizaka}
\affiliation{Nano Electronics Research Laboratories, NEC Corporation,
34 Miyukigaoka, Tsukuba 305-8501, Japan}
\affiliation{INQIE, the University of Tokyo,
4-6-1 Komaba, Meguro-ku, Tokyo 153-8505, Japan}
\author{Tohya Hiroshima}
\affiliation{Nano Electronics Research Laboratories, NEC Corporation,
34 Miyukigaoka, Tsukuba 305-8501, Japan}
\affiliation{
Quantum Computation and Information Project, ERATO-SORST, Japan Science and Technology Agency,\\
Daini Hongo White Building 201, Hongo 5-28-3, Bunkyo-ku, Tokyo 113-0033, Japan}
\date{\today}
%

\begin{abstract}
The scheme of quantum teleportation, where Bob has multiple ($N$)
output ports and obtains the teleported state by simply selecting one of the
$N$ ports, is thoroughly studied. We consider both deterministic version and
probabilistic version of the teleportation scheme aiming to teleport an 
unknown state of a qubit. Moreover, we consider two cases for each version:
(i) the state employed for the teleportation is fixed to a maximally entangled
state, and (ii) the state is also optimized as well as Alice's measurement.
We analytically determine the optimal protocols for all the four cases, and
show the corresponding optimal fidelity or optimal success probability. All
these protocols can achieve the perfect teleportation in the asymptotic limit
of $N\rightarrow\infty$. The entanglement properties of the teleportation
scheme are also discussed.
\end{abstract}

%
\pacs{03.67.Hk, 03.67.Ac, 03.67.Bg, 03.67.Lx}
\maketitle
%
\section{Introduction}
\label{sec: Introduction}

Quantum teleportation \cite{Bennett95b,Boschi98a,Bouwmeester97a,Furusawa98a}
is a fundamental and important protocol for quantum information science and
technology, by which an unknown quantum state is transfered from a sender
(Alice) to a receiver (Bob) exploiting their prior shared entangled state (and
with the assistance of classical communication). In the original (or standard)
teleportation scheme for transferring a state of a qubit
(quantum bit) \cite{Bennett95b}, Alice first performs the Bell-state
measurement on the state $|\chi_{\rm in}\rangle$ to be teleported and her half
of a maximally entangled state
$|\psi^-\rangle_{AB}=(|01\rangle-|10\rangle)/\sqrt{2}$. She then tells the
outcome $k$ to Bob via a classical communication channel. To complete the
teleportation, Bob applies a unitary transformation $\sigma_k$ to his half of
$|\psi^-\rangle_{AB}$, where $\sigma_0\equiv\openone$ and
$(\sigma_1,\sigma_2,\sigma_3)$ are the Pauli matrices. Note that 
continuous-variable teleportation schemes have also been proposed and
intensively studied \cite{Furusawa98a,ContinuousVariable}, where an entangled
state on an infinite dimensional Hilbert space is employed. In this paper,
however, we exclusively consider the schemes with discrete (spin)
variables in a finite dimensional Hilbert space (though we also consider the
limit of infinite dimension).

The quantum teleportation offers a more powerful function than simply
transferring an unknown state \cite{Nielsen97a,Gottesman99a}. Consider that
the state 
$|\varepsilon\rangle=(\openone\otimes\varepsilon)|\psi^-\rangle_{AB}$,
instead of $|\psi^-\rangle_{AB}$, is employed for the standard teleportation
scheme, where $\varepsilon$ is an arbitrary quantum operation. Bob then
obtains  $\sigma_k \varepsilon(\sigma_k|\chi_{\rm in}\rangle)$ as an output of
the teleportation procedure, and thus, obtains
$\varepsilon(|\chi_{\rm in}\rangle)$ when the outcome of the Bell-state
measurement is $k=0$. This implies that the operations of the Bell-state
measurement and the post-selection of the event with $k=0$ (these operations
are denoted by $G$ as a whole) can perform the processing of
$|\chi_{\rm in}\rangle\rightarrow\varepsilon(|\chi_{\rm in}\rangle)$
such that 
$G(|\chi_{\rm in}\rangle\otimes|\varepsilon\rangle)=
\varepsilon(|\chi_{\rm in}\rangle)\otimes|\varepsilon'\rangle$.
The point is that $G$ depends on neither $\varepsilon$ nor
$|\chi_{\rm in}\rangle$, but the fixed $G$ can perform the
manipulation by $\varepsilon$ if an appropriate $|\varepsilon\rangle$ is
provided. The device to manipulate a state in such a way is called a
programmable quantum processor (in short, processor) \cite{Nielsen97a,Gottesman99a,Kim01a,Vidal02c,Fiurasek02a,Fiurasek04a,Brazier05a,Ziman05a,DAriano05a,Hillery06a,Garcia06a},
because the function of the processor is programmed via $|\varepsilon\rangle$.
Moreover, if a processor can be programmed to perform an arbitrary
$\varepsilon$, it is called a universal processor. The standard
teleportation scheme thus offers the function as a universal
processor \cite{Nielsen97a}, because $|\varepsilon\rangle$
is defined for an arbitrary $\varepsilon$ as $|\varepsilon\rangle=(\openone\otimes\varepsilon)|\psi^-\rangle_{AB}$.
Note that, since the form of $|\varepsilon\rangle$ is known
for given $\varepsilon$, we can generate it by various methods, and therefore
an arbitrary state-manipulation can be replaced with a state-preparation as in
Refs. \cite{Gottesman99a,Knill01a}. Note further that, even if 
$|\varepsilon\rangle$ is generated by applying $\varepsilon$ to
$|\psi^-\rangle_{AB}$, we can receive a considerable benefit such that we can
perform $\varepsilon$ before getting an input state $|\chi_{\rm in}\rangle$,
i.e., the time-ordering of these two events can be
exchanged \cite{private,Brukner03a,Imoto04a}.

Unfortunately, however, the universal processor based on the standard
teleportation scheme only works in a probabilistical way. This is because
Bob's unitary transformation $\sigma_k$ with $k\ne0$ generally does not
commute with $\varepsilon$, and hence
$\sigma_k \varepsilon(\sigma_k|\chi_{\rm in}\rangle)\ne
\varepsilon(|\chi_{\rm in}\rangle)$ in general for $k\ne0$.
As a result, the success probability of the universal processor is $1/4$.

\begin{figure}[t]
\centerline{\scalebox{0.44}[0.44]{\includegraphics{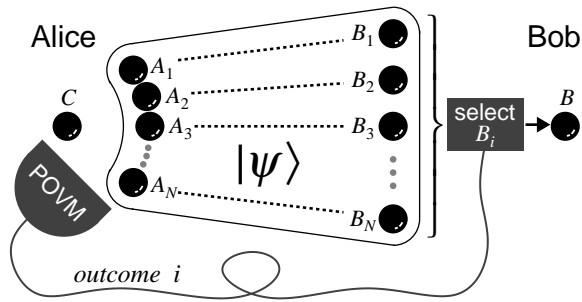}}}
\caption{The setting of the teleportation scheme considered in this paper.
Bob has multiple output ports and obtains the teleported state by simply
selecting one of the $N$ ports according to the outcome ($i$) of Alice's
measurement. To complete the teleportation, no unitary transformation to each
output port is necessary because the state of one of the $N$ ports becomes
the teleported state as it is.}
\label{fig: Setting}
\end{figure}

On the other hand, in the teleportation scheme proposed
by Knill, Laflamme, and Milburn
(KLM) \cite{Knill01a,Franson02a,Grudka08a,Modlawska08a,Kok07a},
Bob has multiple ($N$) output ports and obtains the teleported state by
selecting one of the $N$ ports according to the outcome of Alice's
measurement. To complete the teleportation, Bob further needs to apply a
unitary transformation (phase shift) to the state of the selected port,
as well as the standard teleportation scheme.
As shown in Ref.\ \cite{Ishizaka08b}, however, the teleportation scheme such
that Bob simply selects one of the $N$ ports (without any additional unitary
transformation) is also possible (Fig.\ \ref{fig: Setting}). In fact, the
faithful and deterministic teleportation is asymptotically achieved in the
limit of $N\rightarrow\infty$ \cite{Ishizaka08b}. Let
$|\psi\rangle$ be an entangled state employed for this teleportation scheme
(see Fig.\ \ref{fig: Setting}), and $\varepsilon^{\otimes N}$ denote
the operation of applying $\varepsilon$ to every output port.
Since the operation of simply selecting a port always commutes with
$\varepsilon^{\otimes N}$, if the state
$|\varepsilon\rangle=(\openone\otimes\varepsilon^{\otimes N})|\psi\rangle$
is employed for the teleportation, Bob obtains
$\varepsilon(|\chi_{\rm in}\rangle)$ as an output of the teleportation
procedure, regardless of which port is selected. 
In this way, this teleportation scheme can provide a faithful and deterministic
universal processor in the asymptotic limit of
$N\rightarrow\infty$ \cite{Ishizaka08b}.
Note, however, that such a teleportation scheme must be an approximate and/or
probabilistic one if $N$ is finite, which is a consequence of the no-go
theorem of a faithful and deterministic universal processor with finite
system size \cite{Nielsen97a}.

In this paper, the scheme of quantum teleportation, where Bob simply selects
one of the $N$ ports, is thoroughly studied. We consider both deterministic
version and probabilistic version of the teleportation scheme, and
analytically determine the optimal
protocols. The corresponding optimal fidelity or optimal success probability
are shown as a function of $N$. This paper is organized as follows:
The deterministic (and hence approximate) version of our teleportation scheme
is formulated in Sec.\ \ref{sec: Deterministic version}. The operator $\rho$
defined there [Eq.\ (\ref{eq: rho})] plays an important role for
determining the optimal protocols, and hence the characteristics of $\rho$ is
investigated and summarized in Sec.\ \ref{sec: Characterstics of rho}.
In Sec.\ \ref{sec: Optimal fidelity}, we determine the optimal protocols
of the deterministic version, where we consider two cases:
(i) the state $|\psi\rangle$ employed for the teleportation is fixed to a
maximally entangled state, and (ii) the state $|\psi\rangle$ is also optimized
as well as Alice's measurement. The probabilistic version is then formulated in
Sec.\ \ref{sec: Probabilistic version}, and the optimal protocols are
determined in Sec.\ \ref{sec: Optimal success probability}, where we again
consider two cases (i) and (ii) as in Sec.\ \ref{sec: Optimal fidelity}.
The simplest example of the probabilistic scheme with $N=2$ is explicitly shown
in Sec.\ \ref{sec: Example}. Moreover, the entanglement properties are
discussed in Sec.\ \ref{sec: Entanglement consumption}. In particular, we
focus on the amount of entanglement consumed during the teleportation
procedure. Finally, a summary is given in Sec.\ \ref{sec: Summary}.

%
\section{Deterministic version}
\label{sec: Deterministic version}

In the deterministic version of our teleportation scheme, Bob always accepts
the state of one of the $N$ ports as the teleported state, i.e., the
teleportation is regarded to succeed with unit probability. As mentioned in
the introduction, the deterministic teleportation scheme is necessarily an
approximate one if $N$ is finite.
The optimal protocol is then such that it maximizes the teleportation fidelity
$f$ averaged over all uniformly distributed input pure states. Since the
average fidelity is given by $f=(2F+1)/3$ \cite{Horodecki99a}, the optimal
protocol also maximizes the entanglement fidelity $F$.

Consider that Bob has $N$ qubits: $B_1$, $B_2$, $\cdots$, $B_N$, where each
corresponds to the output port of the teleportation. Alice also has $N$ qubits:
$A_1$, $A_2$, $\cdots$, $A_N$, which are denoted by $A$ as a whole. The state
$|\psi\rangle$ on these $2N$ qubits is employed for teleporting an unknown
state of the $C$ qubit (see Fig.\ \ref{fig: Setting}). Note that the
entanglement fidelity $F$ is maximized when the state employed for the
teleportation is a pure state because of the convexity of $F$. Without loss of
generality, $|\psi\rangle$ can be written as
\begin{equation*}
|\psi\rangle=(O_A\otimes\openone_{B_1\cdots B_N})
|\psi^-\rangle_{A_1B_1}|\psi^-\rangle_{A_2B_2} \cdots
|\psi^-\rangle_{A_NB_N},
\end{equation*}
where
$|\psi^-\rangle=(|01\rangle-|10\rangle)/\sqrt{2}$
is a maximally entangled state (spin-singlet state) and $O$ is an arbitrary
operator that satisfies
$\hbox{tr}O^\dagger O=2^N$ so that $|\psi\rangle$ is normalized.

Alice then performs a joint measurement with $N$ possible outcomes ($1$, $2$,
$\cdots$, $N$) on the $A$ and $C$ qubits. The measurement is described by a
positive operator valued measure (POVM) whose elements are $\{\Pi_i\}$ such
that $\sum_{i=1}^{N}\Pi_i=\openone_{AC}$.
Suppose that she obtains the outcome $i$. She then tells the outcome to Bob
via a classical communication channel, and he discards the $(N-1)$ qubits of
$B_1B_2\cdots B_{i-1}B_{i+1}\cdots B_N$, which are briefly denoted by
${\bar B}_i$. The state of the remaining $B_i$ qubit, which is regarded as the
$B$ qubit, is the teleported state.

The corresponding teleportation channel, which maps the density matrices acting
on the Hilbert space ${\cal H}_C$ to those on ${\cal H}_B$, is thus
\begin{align}
\Lambda(\sigma^{\rm in})
&=
\sum_{i=1}^{N}\left[\hbox{tr}_{A{\bar B_i}C}
\sqrt{\Pi_i}(|\psi\rangle\langle\psi|\otimes\sigma^{\rm in}_{C})
\sqrt{\Pi_i}^\dagger\right]_{B_i\rightarrow B} \nonumber \\
&=
\sum_{i=1}^{N}\hbox{tr}_{AC}
\Pi_i \left(
[(O\otimes\openone)\sigma^{(i)}_{AB}(O^\dagger\otimes\openone)]
\otimes \sigma^{\rm in}_{C}
\right)
\label{eq: channel}
\end{align}
with
\begin{align}
\sigma^{(i)}_{AB}&=\left[\hbox{tr}_{\bar B_i}(
P^-_{A_1B_1}\otimes P^-_{A_2B_2}\otimes \cdots \otimes P^-_{A_NB_N}
)\right]_{B_i\rightarrow B} \nonumber \\
&=\frac{1}{2^{N-1}}P^-_{A_iB} \otimes \openone_{\bar A_i},
\label{eq: sigma}
\end{align}
where $P^-=|\psi^-\rangle\langle\psi^-|$, and ${\bar A}_i$ is a shorthand
notation for $A_1A_2\cdots A_{i-1}A_{i+1}\cdots A_N$.
The entanglement fidelity $F$ for the above channel $\Lambda$ is then given by
\begin{align}
F&=\hbox{tr}P^-_{BD}\left[(\Lambda\otimes\openone)P^-_{CD}\right] \nonumber \\
&=\hbox{tr}\sum_{i=1}^{N} P^-_{BD}
\Pi_{iAC} \left(
[(O\otimes\openone)\sigma^{(i)}_{AB}(O^\dagger\otimes\openone)]
\otimes P^-_{CD}\right) \nonumber \\
&=\frac{1}{2^2}
\sum_{i=1}^{N} \hbox{tr}\Pi_{iAB}
[(O\otimes\openone)\sigma^{(i)}_{AB}(O^\dagger\otimes\openone)] \nonumber \\
&=
\frac{1}{2^2}
\sum_{i=1}^{N} \hbox{tr}\tilde\Pi_{iAB} \sigma^{(i)}_{AB}.
\label{eq: F}
\end{align}
Note that $\Pi_i$ is changed into an operator acting on 
${\cal H}_A\otimes{\cal H}_B$ in the third equality of Eq.\ (\ref{eq: F})
because we used the relationship that 
$(V\otimes\openone)|\psi^-\rangle=(\openone\otimes \sigma_2 V^T \sigma_{2})|\psi^-\rangle$
for any operator $V$, where $\sigma_2$ is the $y$-component of the Pauli
matrices. Moreover, we introduced 
$\tilde\Pi_i=(O^\dagger\otimes\openone)\Pi_i(O\otimes\openone)$ in the last
equality of Eq.\ (\ref{eq: F}), which must satisfy
\begin{equation}
\tilde\Pi_{iAB}\ge 0, \hbox{~~and~~}
\sum_{i=1}^{N}\tilde\Pi_{iAB}=X_A\otimes \openone_B,
\label{eq: POVM}
\end{equation}
where $X=O^\dagger O$ and thus 
\begin{equation}
X\ge0, \hbox{~~and~~} \hbox{tr}X=2^N.
\label{eq: Normalization}
\end{equation}
Hereafter, the subscript of $AB$ in both $\tilde\Pi_i$ and $\sigma^{(i)}$ is
omitted for simplicity.

The optimal protocol is then obtained by maximizing $F$ given by
Eq.\ (\ref{eq: F}) with respect to $\{\tilde\Pi_i\}$ and $X$ under the
constraints of Eqs.\ (\ref{eq: POVM}) and (\ref{eq: Normalization}).
Note that it is possible to consider a more general setting where Alice has
$N_A$ qubits ($N_A\ge N$). This corresponds to consider a $2^{N_A}\times2^{N}$
matrix of $O$ and $2^{N_A+1}\times 2^{N_A+1}$ matrices of $\Pi_i$.
Even in this case, however, $X$ and $\tilde \Pi_i$ to be optimized is a
$2^{N}\times2^{N}$ and $2^{N+1}\times2^{N+1}$ matrix, respectively,
and hence the optimal $X$ and $\tilde \Pi_i$ (and thus the optimal $F$ also)
are not changed even for $N_A>N$.
Therefore, the strategy of employing $N_A>N$ qubits is not helpful for the
purpose of increasing the average fidelity.

For obtaining the optimal protocol of the deterministic version,
and of the probabilistic version also, the operator $\rho$ defined as
\begin{equation}
\rho=\sum_{i=1}^N \sigma^{(i)}
\label{eq: rho}
\end{equation}
plays an important role. Therefore, before discussing the optimal protocols,
we investigate and summarize the characteristics of $\rho$ in the next section.

%
\section{Characteristics of $\rho$}
\label{sec: Characterstics of rho}

Based on the correspondence between qubits and 1/2 spins,
$|0(1)\rangle\leftrightarrow|\frac{1}{2},-\frac{1}{2}(\frac{1}{2})\rangle$,
let us regard each qubit as a 1/2 spin, i.e., $SU(2)$ basis.
The eigenvalues of $\rho$ defined in Eq.\ (\ref{eq: rho}) are given by
\begin{equation}
\lambda_{j}^{-}=\frac{1}{2^N}(\frac{N}{2}-j) \hbox{~~and~~}
\lambda_{j}^{+}=\frac{1}{2^N}(\frac{N}{2}+j+1).
\label{eq: eigenvalues}
\end{equation}
The corresponding eigenstates are
\begin{align} \label{eq:eigenvector}
\lefteqn{|\Psi(\lambda_{j}^{\mp};m) \rangle} \quad\quad\nonumber \\
=&\textstyle |\Phi^{[N]}(j,m+\frac{1}{2},\alpha)\rangle_A |0\rangle_{B} 
 \langle j,m+\frac{1}{2},\frac{1}{2},-\frac{1}{2}|j\pm\frac{1}{2},m\rangle
\nonumber \\
+&\textstyle |\Phi^{[N]}(j,m-\frac{1}{2},\alpha)\rangle_A |1\rangle_{B} 
 \langle j,m-\frac{1}{2},\frac{1}{2},+\frac{1}{2}|j\pm\frac{1}{2},m\rangle, 
\end{align}
where $|\Phi^{[N]}(j,m,\alpha)\rangle=|j,m,\alpha\rangle$ denotes the
orthogonal basis of $N$-spin systems, i.e., the basis of irreducible
representation of $SU(2)^{\otimes N}$. Therefore, $j$ in Eq.\
(\ref{eq: eigenvalues}) represents the spin angular momentum of the $N$-spin
system ($A$ qubits), and hence $j$ runs from $j_{\rm min}$ to $N/2$ where
$j_{\rm min}=0$ $(1/2)$ when $N$ is even (odd). Note that
$|\Psi(\lambda^{\mp}_j;m)\rangle$ are also the eigenstates of the
total spin angular momentum ($s$) of the $(N+1)$-spin system ($A$ and $B$
qubits), and hence $\rho$ is block-diagonal with respect to $s$. The total
spin is given by $s=j\pm1/2$ for $|\Psi(\lambda^{\mp}_j;m)\rangle$, and $m$ in
$|\Psi(\lambda^{\mp}_j;m)\rangle$ runs from $-s$ to $s$.
Note further that $|\Psi(\lambda^{\mp}_j;m)\rangle$ has the implicit
additional degree of freedom with respect to $\alpha$ of $|j,m,\alpha\rangle$,
which takes
$\alpha=1,2,\cdots,g^{[N]}(j)$, where
\begin{equation}
g^{[N]}(j)=\frac{(2j+1)N!}{(N/2-j)!(N/2+1+j)!}.
\label{eq: degeneracy}
\end{equation}
The nonvanishing Clebsch-Gordan (CG) coefficients in
Eq.\ (\ref{eq:eigenvector}) are given by
\begin{align}
{\textstyle \langle j_1,\pm\frac{1}{2},\frac{1}{2},\pm\frac{1}{2}|j_1+\frac{1}{2},\pm1\rangle}& =\sqrt{(j_{1}+{\textstyle \frac{3}{2}})/(2j_{1}+1)},
 \label{eq:CG1} \\
{\textstyle \langle j_{1},\pm\frac{1}{2},\frac{1}{2},\pm\frac{1}{2}|j_1-\frac{1}{2},\pm1\rangle} & =\mp \sqrt{(j_{1}-{\textstyle \frac{1}{2}})/(2j_{1}+1)},
\label{eq:CG2} \\
{\textstyle \langle j_1,\mp\frac{1}{2},\frac{1}{2},\pm\frac{1}{2}|j_1+\frac{1}{2},0\rangle}
& = {\textstyle \mp \langle j_1,\mp\frac{1}{2},\frac{1}{2},\pm\frac{1}{2}|j_1-\frac{1}{2},0\rangle} \nonumber \\ 
& =\sqrt{(j_{1}+{\textstyle \frac{1}{2}})/(2j_{1}+1)}.
\label{eq:CG3}
\end{align}
The proof of the eigenvalue equation
\begin{equation} \label{eq:eigenvalue_equation for N}
\rho| \Psi(\lambda_{j}^{\mp};m) \rangle =
\lambda_{j}^{\mp}|\Psi(\lambda_{j}^{\mp};m)\rangle
\end{equation}
is presented in Appendix \ref{sec: Proof of eigenvalue_equation for N}.

The $N$-spin eigenbasis $|\Phi^{[N]}\rangle $ are obtained recursively;
$|\Phi^{[N-1]}\rangle|\Phi^{[1]}\rangle\rightarrow|\Phi^{[N]}\rangle $,
where $|\Phi^{[N-1]}\rangle$ are $(N-1)$-spin eigenbasis of the first
$(N-1)$ spins ($\bar A_N$ qubits) and $|\Phi^{[1]}\rangle$
are the 1/2-spin state of the $A_N$ qubit. Hence,
$|\Phi^{[N]}(j,\dots)\rangle $ is classified into two;
one is the linear combination of 
$|\Phi^{[N-1]}(j+\frac{1}{2},\dots)\rangle|i\rangle_{A_{N}}$ 
and the other, 
$| \Phi^{[N-1]}(j-\frac{1}{2},\dots) \rangle | i \rangle_{A_{N}}$.
We call the former (latter) are of the type-I (II). Those are given by
\begin{align*}
\lefteqn{|\Phi_{I}^{[N]}(j,m)\rangle} \quad \nonumber \\
=&\textstyle | \Phi^{[N-1]}(j+\frac{1}{2},m+\frac{1}{2}) \rangle 
  |0\rangle_{A_{N}} \langle j+\frac{1}{2},m+\frac{1}{2},\frac{1}{2},-\frac{1}{2}|j,m\rangle  \nonumber \\
+&\textstyle|\Phi^{[N-1]}(j+\frac{1}{2},m-\frac{1}{2})\rangle 
  |1\rangle_{A_{N}} \langle j+\frac{1}{2},m-\frac{1}{2},\frac{1}{2},+\frac{1}{2}|j,m\rangle,
\end{align*}
and
\begin{align*}
\lefteqn{|\Phi_{II}^{[N]}(j,m)\rangle}\quad \nonumber \\
=&\textstyle | \Phi^{[N-1]}(j-\frac{1}{2},m+\frac{1}{2}) \rangle 
  |0\rangle_{A_{N}} \langle j-\frac{1}{2},m+\frac{1}{2},\frac{1}{2},-\frac{1}{2}|j,m\rangle  \nonumber \\
+&\textstyle| \Phi^{[N-1]}(j-\frac{1}{2},m-\frac{1}{2})\rangle 
  |1\rangle_{A_{N}} \langle j-\frac{1}{2},m-\frac{1}{2},\frac{1}{2},+\frac{1}{2}|j,m\rangle.
\end{align*}
According to the different types of $|\Phi^{[N]}\rangle$, eigenstates
$|\Psi(\lambda^{\mp}_j;m)\rangle$ are also classified into two types as
follows: 
\begin{widetext}
\begin{align} \label{eq:Psi_1I}
| \Psi_{\mathrm{I}}(\lambda_{j}^{\mp};m) \rangle 
=&\textstyle 
|\Phi^{[N-1]}(j+\frac{1}{2},m+1,\beta)\rangle_{\bar A_N}|0\rangle_{A_{N}} |0\rangle_{B} 
\langle j,m+\frac{1}{2},\frac{1}{2},-\frac{1}{2}|j\pm\frac{1}{2},m\rangle
\langle j+\frac{1}{2},m+1,\frac{1}{2},-\frac{1}{2}|j,m+\frac{1}{2}\rangle \nonumber \\
+&\textstyle 
|\Phi^{[N-1]}(j+\frac{1}{2},m,\beta)\rangle_{\bar A_N} |1\rangle_{A_{N}} |0\rangle_{B} 
\langle j,m+\frac{1}{2},\frac{1}{2},-\frac{1}{2}|j\pm\frac{1}{2},m\rangle
\langle j+\frac{1}{2},m,\frac{1}{2},\frac{1}{2}|j,m+\frac{1}{2}\rangle \nonumber \\
+&\textstyle 
|\Phi^{[N-1]}(j+\frac{1}{2},m,\beta)\rangle_{\bar A_N} |0\rangle_{A_{N}} |1\rangle_{B} 
\langle j,m-\frac{1}{2},\frac{1}{2},\frac{1}{2}|j\pm\frac{1}{2},m\rangle
\langle j+\frac{1}{2},m,\frac{1}{2},-\frac{1}{2}|j,m-\frac{1}{2}\rangle \nonumber \\
+&\textstyle 
|\Phi^{[N-1]}(j+\frac{1}{2},m-1,\beta)\rangle_{\bar A_N} |1\rangle_{A_{N}} |1\rangle_{B} 
\langle j,m-\frac{1}{2},\frac{1}{2},\frac{1}{2}|j\pm\frac{1}{2},m\rangle
\langle j+\frac{1}{2},m-1,\frac{1}{2},\frac{1}{2}|j,m-\frac{1}{2}\rangle,
\end{align}
and
\begin{align} 
\label{eq:Psi_1II}
| \Psi_{\mathrm{II}}(\lambda_{j}^{\mp};m) \rangle 
=&\textstyle 
|\Phi^{[N-1]}(j-\frac{1}{2},m+1,\beta) \rangle_{\bar A_N} | 0 \rangle_{A_{N}} | 0 \rangle_{B} 
\langle j,m+\frac{1}{2},\frac{1}{2},-\frac{1}{2}|j\pm\frac{1}{2},m\rangle
\langle j-\frac{1}{2},m+1,\frac{1}{2},-\frac{1}{2}|j,m+\frac{1}{2}\rangle \nonumber \\
+&\textstyle 
|\Phi^{[N-1]}(j-\frac{1}{2},m,\beta) \rangle_{\bar A_N} | 1 \rangle_{A_{N}} | 0 \rangle_{B} 
\langle j,m+\frac{1}{2},\frac{1}{2},-\frac{1}{2}|j\pm\frac{1}{2},m\rangle
\langle j-\frac{1}{2},m,\frac{1}{2},\frac{1}{2}|j,m+\frac{1}{2}\rangle \nonumber \\
+&\textstyle 
|\Phi^{[N-1]}(j-\frac{1}{2},m,\beta) \rangle_{\bar A_N} | 0 \rangle_{A_{N}} | 1 \rangle_{B} 
\langle j,m-\frac{1}{2},\frac{1}{2},\frac{1}{2}|j\pm\frac{1}{2},m\rangle
\langle j-\frac{1}{2},m,\frac{1}{2},-\frac{1}{2}|j,m-\frac{1}{2}\rangle \nonumber \\
+&\textstyle 
|\Phi^{[N-1]}(j-\frac{1}{2},m-1,\beta) \rangle_{\bar A_N} | 1 \rangle_{A_{N}} | 1 \rangle_{B} 
\langle j,m-\frac{1}{2},\frac{1}{2},\frac{1}{2}|j\pm\frac{1}{2},m\rangle
\langle j-\frac{1}{2},m-1,\frac{1}{2},\frac{1}{2}|j,m-\frac{1}{2}\rangle.
\end{align}
\end{widetext}
Here, the additional degree of freedom of the $(N-1)$-spin eigenbasis
$|\Phi^{[N-1]}\rangle$ was specified by $\beta$, which takes
$\beta=1,2,\cdots,g^{[N-1]}(j+\frac{1}{2})$ for
$|\Psi_{\mathrm{I}}(\lambda_{j}^{\mp};m)\rangle$ and
$\beta=1,2,\cdots,g^{[N-1]}(j-\frac{1}{2})$ for
$|\Psi_{\mathrm{II}}(\lambda_{j}^{\mp};m)\rangle$.
Here, $g^{[N-1]}(j)$ is given by Eq.\ (\ref{eq: degeneracy}) with
$N\rightarrow(N-1)$.
Note that it is also possible to construct
$|\Psi_{\mathrm{I}}(\lambda_{j}^{\mp};m)\rangle$
and $|\Psi_{\mathrm{II}}(\lambda_{j}^{\mp};m)\rangle$
by using the $(N-1)$-spin eigenbasis for the $\bar A_i$ qubits 
(instead of the $\bar A_N$ qubits) and the states of the $A_iB$ qubits. Let us
denote the resultant $(N-1)$-spin eigenbasis by
$|\Phi^{[N-1]\prime}(j,m,\beta')\rangle$, which are unitarily equivalent to
$|\Phi^{[N-1]}(j,m,\beta)\rangle$. The unitary transformation depends only
on $\beta$ and $\beta'$ for each $j$ \cite{Messiah}. Namely,
\begin{equation}
| \Phi^{[N-1]\prime}(j,m,\beta') \rangle 
=\sum_{\beta} \big[ U(j) \big]_{\beta' \beta} |\Phi^{[N-1]}(j,m,\beta)\rangle
\label{eq: unitary equivalent}
\end{equation}
holds with $U(j)$ being a unitary matrix.

As mentioned above, $\rho$ is block-diagonal with respect to the total spin
angular momentum $s$, and let us denote the block-matrices by $\rho(s)$.
Since $j=s\mp1/2$ for $\lambda^{\mp}_j$, $\rho(s)$ is written as
$\rho(s)=\rho_-(s)\oplus\rho_+(s)$ with
\begin{equation}
\rho_{\mp}(s)=
\lambda^{\mp}_{s\mp1/2} \sum_{m=-s}^s\sum_{\alpha}
|\Psi(\lambda_{s\mp1/2}^{\mp};m)\rangle
\langle\Psi(\lambda_{s\mp1/2}^{\mp};m)|,
\label{eq: rho +-}
\end{equation}
or equivalently,
\begin{align*}
\rho_{\mp}(s)
&=\lambda^{\mp}_{s\mp1/2} \sum_{m=-s}^s\Big[\sum_{\beta}
|\Psi_{\mathrm{I}}(\lambda_{s\mp1/2}^{\mp};m)\rangle
\langle\Psi_{\mathrm{I}}(\lambda_{s\mp1/2}^{\mp};m)| \\
&\quad\quad\quad\quad+\sum_{\beta}
|\Psi_{\mathrm{II}}(\lambda_{s\mp1/2}^{\mp};m)\rangle
\langle\Psi_{\mathrm{II}}(\lambda_{s\mp1/2}^{\mp};m)|\Big].
\end{align*}
The degeneracy of $\lambda^{-}_{s-1/2}=(N/2+1/2-s)/2^N$ is
\begin{equation}
(2s+1)\Big\{g^{[N-1]}(s)+g^{[N-1]}(s-1)\Big\}
=(2s+1)g^{[N]}(s-\textstyle\frac{1}{2}),
\label{eq: degeneracy -}
\end{equation}
where the first and the second term on the left hand side originates from 
$|\Psi_{\mathrm{I}}(\lambda_{s-1/2}^{-};m)\rangle$ and
$|\Psi_{\mathrm{II}}(\lambda_{s-1/2}^{-};m)\rangle$, respectively.
From Eqs.\ (\ref{eq:Psi_1I}), (\ref{eq:Psi_1II}), and
(\ref{eq: unitary equivalent}), and using the explicit form of the
CG coefficients [Eqs.\ (\ref{eq:CG1})-(\ref{eq:CG3})],
it is found that
\begin{align}
\label{eq: overlap 1}
\lefteqn{\langle\psi^{-}_{A_i B}|\Psi_{\mathrm{I}}(\lambda_{s-1/2}^{-};m)\rangle} \quad\nonumber \\
&=\frac{(s-m)+(s+m)}{\sqrt{2}\sqrt{2s(2s+1)}}
|\Phi^{[N-1]\prime}(s,m,\beta')\rangle_{\bar A_i} \nonumber \\
&=\sqrt{\frac{s}{2s+1}}\sum_{\beta}\big[ U(s) \big]_{\beta' \beta} |\Phi^{[N-1]}(s,m,\beta)\rangle_{\bar A_i},
\end{align}
and
\begin{equation}
\label{eq: overlap 2}
\langle\psi^{-}_{A_i B}|\Psi_{\mathrm{II}}(\lambda_{s-1/2}^{-};m)\rangle=0.
\end{equation}
Likewise, the degeneracy of $\lambda^{+}_{s+1/2}=(N/2+3/2+s)/2^N$ is
\begin{equation}
(2s+1)\Big\{g^{[N-1]}(s+1)+g^{[N-1]}(s)\Big\}
=(2s+1)g^{[N]}(s+\textstyle\frac{1}{2}),
\label{eq: degeneracy +}
\end{equation}
where the first and the second term on the left hand side originates from 
$|\Psi_{\mathrm{I}}(\lambda_{s+1/2}^{+};m)\rangle$ and
$|\Psi_{\mathrm{II}}(\lambda_{s+1/2}^{+};m)\rangle$, respectively.
Moreover, we have
\begin{equation}
\label{eq: overlap 3}
\langle\psi^{-}_{A_i B}|\Psi_{\mathrm{I}}(\lambda_{s+1/2}^{+};m)\rangle=0,
\end{equation}
and
\begin{align}
\label{eq: overlap 4}
\lefteqn{\langle\psi^{-}_{A_i B}|\Psi_{\mathrm{II}}(\lambda_{s+1/2}^{+};m)\rangle} \quad\nonumber \\
&=-
\frac{(s+m+1)+(s-m+1)}{\sqrt{2}\sqrt{2(s+1)(2s+1)}}
|\Phi^{[N-1]\prime}(s,m,\beta')\rangle_{\bar A_i} \nonumber \\
&=-\sqrt{\frac{s+1}{2s+1}}\sum_{\beta}\big[ U(s) \big]_{\beta' \beta} |\Phi^{[N-1]}(s,m,\beta)\rangle_{\bar A_i}.
\end{align}

Let us now introduce the states of
\begin{equation}
|\xi^{(i)}(s,m,\beta)\rangle=
|\psi^{-}\rangle_{A_{i}B}|\Phi^{[N-1]}(s,m,\beta)\rangle_{\bar A_i}.
\end{equation}
Using these states, $\sigma^{(i)}$ of Eq.\ (\ref{eq: sigma}) is written as
\begin{equation*}
\sigma^{(i)}= \sum_{s=s_{\rm min}}^{(N-1)/2} \sigma^{(i)}(s)
\end{equation*}
with
\begin{equation*}
\sigma^{(i)}(s)
=\frac{1}{2^{N-1}} 
\sum_{m=-s}^s\sum_{\beta} | \xi^{(i)}(s,m,\beta) \rangle\langle \xi^{(i)}(s,m,\beta) |,
\end{equation*}
where $s_{\rm min}=0$ $(1/2)$ when $N-1$ is even (odd). 
From Eqs.\ (\ref{eq: overlap 1}), (\ref{eq: overlap 2}),
(\ref{eq: overlap 3}) and (\ref{eq: overlap 4}), and noting the orthogonality
of
$\sum_{\gamma} \big[U(s)\big]_{\gamma\beta}\big[U(s)\big]_{\gamma\beta'}^*=\delta_{\beta,\beta^{\prime}}$, we have
\begin{align}
\lefteqn{\langle\xi^{(i)}(s,m,\beta)|\rho(s'')^{-1/y}
|\xi^{(i)}(s',m',\beta')\rangle} \quad\quad\quad\quad\quad\quad\quad\quad\quad
\nonumber \\
=&\delta_{s,s''}\delta_{s',s''}\delta_{m,m'}\delta_{\beta,\beta'} c(s,y),
\label{eq:matrix_element}
\end{align}
where $y$ is arbitrary real, and
\begin{equation}
c(s,y)=\frac{s}{2s+1}\Big(\lambda^{-}_{s-1/2}\Big)^{-1/y}+
\frac{s+1}{2s+1}\Big(\lambda^{+}_{s+1/2}\Big)^{-1/y}.
\label{eq:c}
\end{equation}
Note that $c(s,y)$ depends only on $s$ (for a fixed $y$ and $N$). As a result,
it is found that both $\rho$ and $\sigma^{(i)}$ are simultaneously 
block-diagonal with respect to $s$, and hence the block-matrices $\rho(s)$ and
$\sigma^{(i)}(s')$ are orthogonal to each other for $s\ne s'$.

%
\section{Optimal fidelity}
\label{sec: Optimal fidelity}

%
\subsection{Maximally entangled $|\psi\rangle$}

\begin{figure}[t]
\centerline{\scalebox{0.45}[0.45]{\includegraphics{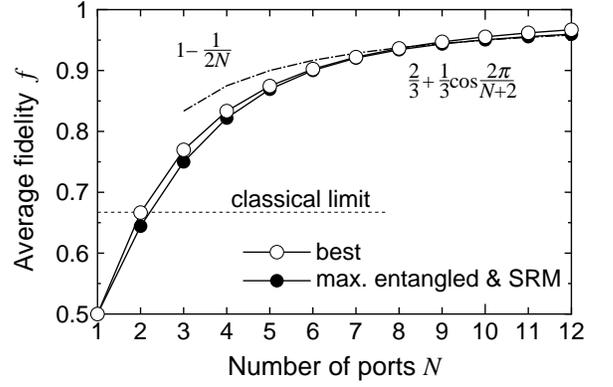}}}
\caption{The average fidelity ($f$) in the deterministic scheme as 
a function of number of output ports ($N$). The asymptotic behavior
[$1-1/(2N)$] in the case of the maximally entangled $|\psi\rangle$ is
also plotted by a dash-dot line.
}
\label{fig: fidelity}
\end{figure}

Let us first consider the case where the state $|\psi\rangle$ employed for
the deterministic teleportation is fixed to
a maximally entangled state, i.e., $|\psi\rangle=|\psi^-\rangle^{\otimes N}$.
This corresponds to the case where $X$ is fixed to $X=O^\dagger O=\openone$,
and only the measurement performed by Alice is optimized to
maximize the average fidelity $f$. As shown in Ref.\ \cite{Ishizaka08b}, the
optimal measurement is the square-root measurement (SRM)
(also known as a pretty good measurement or least-squares measurement)
\cite{Hausladen94a,Hausladen96a,Ban97a,Sasaki98a,Kato03a,Eldar04a}.
The optimal POVM elements are thus
\begin{equation}
\Pi_i=\rho^{-1/2}\sigma^{(i)}\rho^{-1/2},
\label{eq: SRM}
\end{equation}
where $\rho^{-1}$ is defined on the support of $\rho$, and
we implicitly assume that the excess term
\begin{equation}
\Delta=\frac{1}{N}(\openone-\sum_{i=1}^N \rho^{-1/2}\sigma^{(i)}\rho^{-1/2})
\end{equation}
is added to every $\Pi_i$ so that the POVM elements sum to identity.
Note that $\hbox{tr}\sigma^{(i)}\Delta=0$.
From Eqs.\ (\ref{eq:matrix_element}) and (\ref{eq:c}), the optimal entanglement
fidelity is calculated as
\begin{align}
F&=\frac{1}{2^2}\hbox{tr}\sum_{s=s_{\rm min}}^{(N-1)/2} \sum_{i=1}^N 
\rho(s)^{-1/2} \sigma^{(i)}(s) \rho(s)^{-1/2} \sigma^{(i)}(s) \nonumber \\
&=\frac{N}{2^{2N}}\sum_{s=s_{\rm min}}^{(N-1)/2}
(2s+1)g^{[N-1]}(s)c(s,2)^2 \nonumber \\
&=\frac{1}{2^{N+3}}\sum_{k=0}^{N}
\bigg(\frac{N-2k-1}{\sqrt{k+1}}+\frac{N-2k+1}{\sqrt{N-k+1}}\bigg)^2
\binom{N}{k}.
\label{eq: Optimal fidelity}
\end{align}
The corresponding average fidelity $f$ as a function of $N$ is plotted by
closed circles in Fig.\ \ref{fig: fidelity}. For $N\ge3$, the fidelity exceeds
the classical limit $f_{\rm cl}=2/3$, which is the best fidelity via a
classical channel only \cite{Horodecki99a}.
For $N\rightarrow\infty$, we find that $F\rightarrow 1-3/(4N)$, and hence
\begin{equation}
f\rightarrow 1-1/(2N) \hbox{~~~for $N\rightarrow\infty$}.
\label{eq: Asymptotic optimal fidelity}
\end{equation}

The above protocol of employing maximally entangled $|\psi\rangle$ and SRM
can be easily extended to the case of teleporting an unknown state of a qudit
($d$-dimensional system), where $|\psi\rangle=|\phi^+\rangle^{\otimes N}$ with 
$|\phi^+\rangle=(1/\sqrt{d})\sum_{k=0}^{d-1}|kk\rangle$,
and the POVM elements are given by Eq.\ (\ref{eq: SRM}) with
$\sigma^{(i)}=(1/d^{N-1}) P_{A_iB}^+\otimes\openone_{\bar A_i}$,
where $P^+=|\phi^+\rangle\langle\phi^+|$.
As mentioned in Ref.\ \cite{Ishizaka08b}, the average fidelity is lower
bounded as
\begin{equation}
f\ge 1- d(d-1)/N.
\label{eq: Bounded fidelity}
\end{equation}
The proof is presented in Appendix \ref{sec: Proof of bounded fidelity}.

To investigate the property of the teleportation channel $\Lambda$ 
[Eq.\ (\ref{eq: channel}) extended to the qudit case], let us
consider the state isomorphic to the channel:
$\omega_{BD}=(\Lambda\otimes\openone)P^+_{CD}$. Using
$(U\otimes\openone)|\phi^+\rangle=(\openone\otimes U^T)|\phi^+\rangle$,
we have for $O=\openone$
\begin{align*}
\lefteqn{(U_B\otimes U_{D}^*)\omega_{BD}(U_{B}^\dagger \otimes U_{D}^T)}
\quad \\
&=\sum_{i=1}^{N} \hbox{tr}_{AC}
(U_{A_i}^*\otimes U_{C})\Pi_{iAC}(U_{A_i}^T \otimes U_{C}^\dagger)
(\sigma^{(i)}_{AB}\otimes P^+_{CD}) \\
&=\sum_{i=1}^{N} \hbox{tr}_{AC}
(U_{A}^*\otimes U_{C})\Pi_{iAC}(U_{A}^T \otimes U_{C}^\dagger)
(\sigma^{(i)}_{AB}\otimes P^+_{CD}),
\end{align*}
where $U_A$ denotes $U_{A_1}\otimes U_{A_2}\otimes \cdots \otimes U_{A_N}$,
and $\sigma^{(i)}_{AB}=U_{\bar A_i}^T \sigma^{(i)}_{AB} U_{\bar A_i}^*$ was
used in the second equality. Since $\sigma^{(i)}_{AC}$ (and thus $\rho_{AC}$)
is invariant under the $(U_{A}^*\otimes U_C)$-twirling, $\{\Pi_{iAC}\}$ of
SRM (and $\Delta_{AC}$) is also invariant under the twirling. As a result,
$\omega_{BD}$ is invariant under the $(U_B\otimes U_{D}^*)$-twirling. This
implies that $\omega_{BD}$ is an isotropic state, and therefore the
teleportation channel is a depolarizing channel.

%
\subsection{Optimal $|\psi\rangle$} 
\label{subsec: Optimal state in the deterministic version}

Let us next consider the case where both $|\psi\rangle$ and Alice's
measurement are optimized. The optimal POVM elements are
\begin{equation}
\tilde \Pi_i=\sum_{s=s_{\rm min}}^{(N-1)/2} z(s) \rho(s)^{-1/y(s)}
\sigma^{(i)}(s)\rho(s)^{-1/y(s)},
\label{eq: Best deterministic measurement}
\end{equation}
where $y(s)$ is defined through
\begin{equation}
\bigg(\frac{\lambda_{s-1/2}^-}{\lambda_{s+1/2}^+}\bigg)^{1/y(s)}\frac{s+1}{s}
=\sin\frac{2\pi(s+1)}{N+2}/\sin\frac{2\pi s}{N+2}
\equiv D(s),
\label{eq: D(s)}
\end{equation}
and $z(s)$ is given by
\begin{equation*}
z(s)=\frac{2^{N+1}\big(\lambda_{s-1/2}^-\big)^{2/y(s)-1}}
{(N+2)sg^{[N]}(s-1/2)}
\sin^2\frac{2\pi s}{N+2}.
\end{equation*}
Note that the form of the optimal $\tilde\Pi_i$ resembles the form of SRM, but
$y(s)$ is not generally equal to 2; it is a function of $s$ (and $N$). In this
way, the optimal measurement becomes, say, the generalized SRM if both
$|\psi\rangle$ and Alice's measurement are optimized. Note further that we
implicitly assume that the excess term $\tilde\Delta$ is
added to every $\tilde\Pi_i$, as in the case of SRM, so that the POVM elements
sum to $X\otimes\openone$ [see Eq.\ (\ref{eq: POVM})].
The optimal state $|\psi\rangle$ is specified through $X$ given by
\begin{equation}
X=\sum_{j=j_{\rm min}}^{N/2}\gamma(j) \openone(j)_A,
\label{eq: deterministic X}
\end{equation}
where $\openone(j)_A$ is the identity on the subspace spanned by
$|\Phi^{[N]}(j,\cdots)\rangle_A$, and
\begin{equation}
\gamma(j)= \frac{2^{N+2}}{(N+2)(2j+1)g^{[N]}(j)}
\sin^2\frac{\pi (2j+1)}{N+2}.
\end{equation}
For the above choice of $\{\tilde\Pi_i\}$,
\begin{align*}
\sum_{i=1}^N\tilde\Pi_i
=&\sum_{s=s_{\rm min}}^{(N-1)/2} z(s) \rho(s)^{1-2/y(s)} \\
=&z(s_{\rm min})\big(\lambda_{s_{\rm min}-1/2}^-\big)^{1-2/y(s_{\rm min})}
\openone_-(s_{\rm min})\\
&+\sum_{s=s_{\rm min}+1}^{(N-1)/2}
\Big\{z(s)\big(\lambda_{s-1/2}^-\big)^{1-2/y(s)}\openone_-(s) \\
&\quad +z(s-1)\big(\lambda_{s-1/2}^+\big)^{1-2/y(s-1)}\openone_+(s-1) \Big\} \\
&+z({\textstyle\frac{N-1}{2}})
\big(\lambda_{N/2}^+\big)^{1-2/y({\textstyle\frac{N-1}{2}})}
\openone_+({\textstyle \frac{N-1}{2}}),
\end{align*}
where 
\begin{equation}
\openone_\mp(s)=
\sum_{m=-s}^s\sum_{\alpha}
|\Psi(\lambda_{s\mp1/2}^{\mp};m)\rangle
\langle\Psi(\lambda_{s\mp1/2}^{\mp};m)|
\label{eq: identity +-}
\end{equation}
is the identity on the support of $\rho_{\mp}(s)$
[see Eq.\ (\ref{eq: rho +-})]. Using
\begin{align}
z(s)\big(\lambda_{s-1/2}^-\big)^{1-2/y(s)}
&=z(s-1)\big(\lambda_{s-1/2}^+\big)^{1-2/y(s-1)} \nonumber \\
&=\gamma(s-1/2),
\end{align}
and the following relation:
\begin{equation}
\openone_-(s)+\openone_+(s-1)=\openone(s-{\textstyle \frac{1}{2}})_A\otimes\openone_B
\label{eq: relation of identity}
\end{equation}
obtained from Eq.\ (\ref{eq:eigenvector}) and the CG coefficients,
we have
\begin{align}
\sum_{i=1}^N\tilde\Pi_i
=&\sum_{j=j_{\rm min}}^{N/2-1} \gamma(j) \openone(j)_A\otimes\openone_B
+\gamma({\textstyle \frac{N}{2}})\openone_+({\textstyle\frac{N-1}{2}})
\nonumber \\
\le & X_A\otimes \openone_B.
\label{eq: fulfillment of XotimesI}
\end{align}
Therefore, the constraint of Eq.\ (\ref{eq: POVM}) can be satisfied for
an appropriate choice of $\tilde\Delta\ge0$. Moreover, since
\begin{equation}
\hbox{tr}X=\frac{2^{N+2}}{N+2}\sum_{j=j_{\rm min}}^{N/2}
\sin^2\frac{\pi (2j+1)}{N+2}=2^N,
\end{equation}
the constraint of Eq.\ (\ref{eq: Normalization}) is also satisfied.

The optimal entanglement fidelity is then calculated as
\begin{align}
F=&\frac{1}{2^{N+1}}\hbox{tr}\sum_{s=s_{\rm min}}^{(N-1)/2}
z(s) c(s,y(s)) \rho(s)^{1-1/y(s)} \nonumber \\
=&\frac{1}{(N+2)}\sum_{s=s_{\rm min}}^{(N-1)/2}
\frac{\sin^2\frac{2\pi s}{N+2}}{sg^{[N]}(s-1/2)}\Big[D(s)+1\Big] \nonumber \\
&\times \Big[D(s)\frac{s^2}{s+1}
\frac{\lambda_{s+1/2}^+}{\lambda_{s-1/2}^-}g^{[N]}(s+1/2)
+sg^{[N]}(s-1/2)  \Big] \nonumber \\
=&\frac{1}{(N+2)}\sum_{s=s_{\rm min}}^{(N-1)/2}
\sin^2\frac{2\pi s}{N+2} \Big[D(s)+1\Big]^2 \nonumber \\
=&\cos^2\frac{\pi}{N+2},
\label{eq: Best entanglement fidelity}
\end{align}
where $D(s)$ has been defined in Eq.\ (\ref{eq: D(s)}). The corresponding
average fidelity
\begin{equation}
f=\frac{2}{3}+\frac{1}{3}\cos\frac{2\pi}{N+2}
\label{eq: Best fidelity}
\end{equation}
is plotted by open circles in Fig.\ \ref{fig: fidelity}. 
The optimality of Eq.\ (\ref{eq: Best entanglement fidelity}) is proved in
Appendix \ref{sec: Best fidelity}. Since both
$|\psi\rangle$ and Alice's measurement are simultaneously optimized, this
is the best fidelity in the teleportation scheme such that Bob simply selects
one of the multiple qubits.
It is found from the figure that the best fidelity is nearly achieved by the
protocol of employing maximally entangled $|\psi\rangle$ and SRM.
Note, however, that the asymptotic behaviors of the fidelity are different
from each other: $f\rightarrow 1-{\cal O}(1/N)$ for maximally entangled
$|\psi\rangle$ [Eq.\ (\ref{eq: Asymptotic optimal fidelity})], while 
$f\rightarrow 1-{\cal O}(1/N^2)$ if $|\psi\rangle$ is also optimized
[Eq.\ (\ref{eq: Best fidelity})].

%
\section{Probabilistic version}
\label{sec: Probabilistic version}

In the probabilistic scheme, the teleportation sometimes fails, but if
the teleportation succeeds, the state is faithfully teleported with perfect
fidelity $f=1$. The optimal protocol is then such that it maximizes the
average success probability.

Let $\{\Pi_0,\Pi_1,\Pi_2,\cdots,\Pi_N\}$ be the POVM elements of Alice's
measurement. Suppose that the teleportation fails if $\Pi_0$ is obtained in
her measurement; otherwise, when $\Pi_i$ with $i\ne0$ is obtained, the
teleportation faithfully succeeds, where the state of the $B_i$ qubit is
exactly equal to the input state of the $C$ qubit
(see Fig.\ \ref{fig: Setting}). As in the case of the deterministic version
discussed in Sec.\ \ref{sec: Deterministic version}, the teleportation channel
is given by Eq.\ (\ref{eq: channel}) (when the teleportation succeeds).
However, the channel is trace-nonpreserving in this case, and
\begin{equation}
\hbox{tr}\Lambda(\sigma^{\rm in})
=\frac{1}{2^N}
\sum_{i=1}^{N}\hbox{tr}_{AC}
\Pi_{iAC} \left( OO^\dagger\otimes \sigma^{\rm in}_{C} \right)
\end{equation}
corresponds to the success probability (when the input state is
$\sigma^{\rm in}$). The success probability $p$ averaged over all uniformly
distributed input pure states is then given by
\begin{equation}
p=\frac{1}{2^{N}}\sum_{i=1}^{N}\hbox{tr} \Pi_i (OO^\dagger\otimes\frac{\openone_C}{2})
=\frac{1}{2^{N+1}}\sum_{i=1}^{N}\hbox{tr} \tilde \Pi_i,
\label{eq: p}
\end{equation}
where we again introduced 
$\tilde\Pi_i=(O^\dagger\otimes\openone)\Pi_i(O\otimes\openone)$.
Note that $p$ agrees with the success probability when half of $P^-_{CD}$
is teleported as in the entanglement swapping. The entanglement fidelity is
thus given by
\begin{equation*}
F=\frac{1}{p}\hbox{tr}P^-_{BD}\left[(\Lambda\otimes\openone)P^-_{CD}\right]
=\frac{1}{2^2p}
\sum_{i=1}^{N} \hbox{tr} \tilde \Pi_{iAB} \sigma^{(i)}_{AB}.
\end{equation*}
Since $F=1$ for the faithful teleportation, it is found that
$\hbox{tr}\tilde \Pi_i (\openone-P^-)_{A_iB}=0$ must hold for $i=1,2,\cdots,N$.
This implies that $\tilde\Pi_i$ must have the form of 
\begin{equation}
\tilde \Pi_i =P^-_{A_iB}\otimes \tilde\Theta_{i\bar A_i}
\hbox{~~for $i=1,2,\cdots,N$},
\end{equation}
where $\{\tilde \Theta_i\}$ with $i=1,2,\cdots,N$ must satisfy
\begin{equation}
\tilde\Theta_i \ge 0 \hbox{~~~and~~~}
\sum_{i=1}^N P^-_{A_iB}\otimes\tilde\Theta_{i\bar A_i}\le X_A\otimes\openone_B,
\label{eq: Theta}
\end{equation}
because $\Pi_i\ge0$ and $\sum_{i=1}^N\Pi_i\le\openone$.
Here, we again introduced $X=O^\dagger O$, which must satisfy
Eq.\ (\ref{eq: Normalization}). The average success probability is then written as
\begin{equation}
p=\frac{1}{2^{N+1}}\sum_{i=1}^{N} \hbox{tr} \tilde \Theta_{i\bar A_i}.
\label{eq: p2}
\end{equation}
Therefore, the optimal protocol of the probabilistic version is obtained by
maximizing $p$ given by Eq.\ (\ref{eq: p2}) with respect to
$\{\tilde\Theta_i\}$ and $X$ under the constraints of Eqs.\ (\ref{eq: Theta})
and (\ref{eq: Normalization}).

%
\section{Optimal success probability}
\label{sec: Optimal success probability}

%
\subsection{Maximally entangled $|\psi\rangle$}
\label{subsec: Maximally entangled state in the probabilistic version}

\begin{figure}[t]
\centerline{\scalebox{0.45}[0.45]{\includegraphics{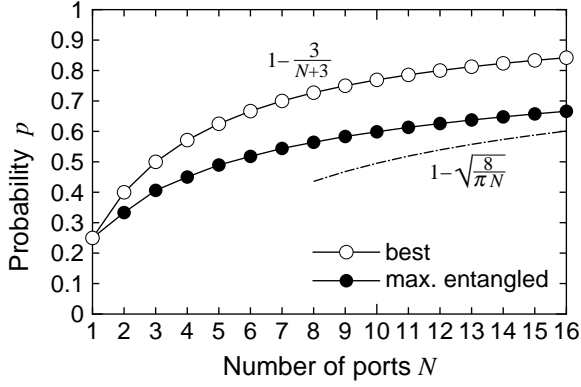}}}
\caption{The average success probability ($p$) in the probabilistic scheme
as a function of number of output ports ($N$). The asymptotic behavior
[$1-\sqrt{8/(\pi N)}$] in the case of the maximally entangled $|\psi\rangle$
is also plotted by a dash-dot line.
}
\label{fig: probability}
\end{figure}

Let us first consider the case where the state $|\psi\rangle$ is fixed as
$|\psi\rangle=|\psi^-\rangle^{\otimes N}$, i.e., $X=O^\dagger O=\openone$,
and only the measurement performed by Alice is optimized to maximize the
success probability $p$. The optimal POVM elements are given by
\begin{equation}
\tilde\Theta_{i\bar A_i}=\frac{1}{2^{N-1}}\sum_{s=s_{\rm min}}^{(N-1)/2}
\frac{1}{\lambda_{s+1/2}^+}\openone(s)_{\bar A_i},
\label{eq: Optimal probabilistic measurement}
\end{equation}
where $\openone(s)_{\bar A_i}$ is the identity on the subspace spanned by
$|\Phi^{[N-1]}(s,\cdots)\rangle_{\bar A_i}$. For this choice,
\begin{align}
\sum_{i=1}^N P^-_{A_iB}\otimes \tilde\Theta_{i\bar A_i}&=
\sum_{s=s_{\rm min}}^{(N-1)/2} \frac{1}{\lambda_{s+1/2}^+}
\sum_{i=1}^N \frac{1}{2^{N-1}} P^-_{A_iB}\otimes \openone(s)_{\bar A_i}
\nonumber \\
&=\sum_{s=s_{\rm min}}^{(N-1)/2} \frac{1}{\lambda_{s+1/2}^+}\rho(s)\le\openone,
\end{align}
because $\lambda_{s+1/2}^+$ is the largest eigenvalue of $\rho(s)$,
and hence the constraint of Eq.\ (\ref{eq: Theta}) is satisfied.
The optimal success probability is then calculated as
\begin{align}
p&=\frac{1}{2^{2N}} \sum_{s=s_{\rm min}}^{(N-1)/2}
\frac{N}{\lambda_{s+1/2}^+}\hbox{tr}\openone(s)_{\bar A_i} \nonumber \\
&=\frac{1}{2^{2N}} \sum_{s=s_{\rm min}}^{(N-1)/2}
\frac{N(2s+1)g^{[N-1]}(s)}{\lambda_{s+1/2}^+} \nonumber \\
&=\frac{1}{2^N} \sum_{s=s_{\rm min}}^{(N-1)/2}
\frac{(2s+1)^2 N!}{(\frac{N-1}{2}-s)!(\frac{N+3}{2}+s)!},
\label{eq: Optimal probability}
\end{align}
which is plotted by closed circles in Fig.\ \ref{fig: probability}.
Moreover, we find
\begin{align}
p\rightarrow 1-\sqrt{8/(\pi N)} \hbox{~~~for $N\rightarrow\infty$},
\label{eq: Asymptotic optimal probability}
\end{align}
and therefore, this protocol achieves the unit success probability
in the asymptotic limit of $N\rightarrow\infty$.
The optimality of Eq.\ (\ref{eq: Optimal probability}) is proved in
Appendix \ref{sec: Optimal probability}.

%
\subsection{Optimal $|\psi\rangle$}
\label{subsec: Optimal state in the probabilistic version}

Let us next consider the case where both $|\psi\rangle$ and
Alice's measurement are optimized simultaneously.
The optimal POVM elements are given by
\begin{equation}
\tilde\Theta_{i\bar A_i}=\sum_{s=s_{\rm min}}^{(N-1)/2}
u(s)\openone(s)_{\bar A_i},
\label{eq: Best probabilistic measurement}
\end{equation}
where
\begin{equation}
u(s)=\frac{2^{N+1}h(N)(2s+1)}{Ng^{[N-1]}(s)}
\end{equation}
with $h(N)=6/[(N+1)(N+2)(N+3)]$. The optimal state $|\psi\rangle$ is specified
through $X$ given by
\begin{equation}
X=\sum_{j=j_{\rm min}}^{N/2}\nu(j) \openone(j)_A
\hbox{~with~} \nu(j)= \frac{2^{N}h(N)(2j+1)}{g^{[N]}(j)}.
\label{eq: probabilistic X}
\end{equation}
For the above choice,
\begin{align*}
\lefteqn{\frac{1}{2^{N-1}}
\sum_{i=1}^N P^-_{A_iB}\otimes \tilde\Theta_{i\bar A_i}
=\sum_{s=s_{\rm min}}^{(N-1)/2} u(s)\rho(s)} \quad\quad\\
=&u(s_{\rm min})\lambda_{s_{\rm min}-1/2}^-\openone_-(s_{\rm min})\\
+&\sum_{s=s_{\rm min}+1}^{(N-1)/2}
\Big\{u(s)\lambda_{s-1/2}^-\openone_-(s) \\
+&u(s-1)\lambda_{s-1/2}^+\openone_+(s-1) \Big\}
+u({\textstyle\frac{N-1}{2}})
\lambda_{N/2}^+\openone_+({\textstyle \frac{N-1}{2}}),
\end{align*}
where $\openone_\mp(s)$ is given by Eq.\ (\ref{eq: identity +-}).
Using 
\begin{equation*}
u(s)\lambda_{s-1/2}=u(s-1)\lambda_{s-1/2}=\nu(s-1/2)/2^{N-1},
\end{equation*}
and Eq.\ (\ref{eq: relation of identity}), the fulfillment of
Eq.\ (\ref{eq: Theta}) is confirmed in the same way as
Eq.\ (\ref{eq: fulfillment of XotimesI}). The constraint of
Eq.\ (\ref{eq: Normalization}) is also satisfied because
\begin{equation}
\hbox{tr}X=2^N h(N)\sum_{j=j_{\rm min}}^{N/2}
(2j+1)^2=2^N.
\end{equation}
The optimal success probability is then
\begin{align}
p&=\frac{N}{2^{N+1}} \sum_{s=s_{\rm min}}^{(N-1)/2}
u(s)\hbox{tr}\openone(s)_{\bar A_i} \nonumber \\
&=h(N)\sum_{s=s_{\rm min}}^{(N-1)/2}(2s+1)^2
=\frac{N}{N+3}=1-\frac{3}{N+3},
\label{eq: Best probability}
\end{align}
which is plotted by open circles in Fig.\ \ref{fig: probability}.
The optimality of Eq.\ (\ref{eq: Best probability}) is proved in
Appendix \ref{sec: Best probability}. Here, let us recall that the success
probability in the KLM scheme is equal to $p=1-1/(N+1)$ \cite{Knill01a}.
Comparing this and
Eq.\ (\ref{eq: Best probability}), it is found that the number of ports
$N$ in our scheme must be just three times larger than that of the KLM scheme
to achieve the same success probability. Therefore, this three times
increase of the number of ports is, in some sense, regarded as the cost we
have to pay to remove Bob's unitary transformation.

It has been shown in Ref.\ \cite{Grudka08a} that the success probability of the
(probabilistic) KLM scheme is maximized when a maximally entangled state is
employed. On the other hand, in contrast to the KLM scheme, 
the success probability in our scheme is considerably enhanced by optimizing
$|\psi\rangle$ as shown in Fig.\ \ref{fig: probability}.
This implies that non-maximally entangled $|\psi\rangle$ can
provide considerably larger success probability than that of a maximally
entangled $|\psi\rangle$. Interestingly,
we have from Eq.\ (\ref{eq: probabilistic X})
\begin{align*}
\sigma_{B_i}&=
\hbox{tr}_{A \bar B_i}|\psi\rangle\langle\psi|=\frac{1}{2^N}
\sigma_{2}\big(\hbox{tr}_{\bar B_i}X^{T}_{B_1\cdots B_N}\big)\sigma_{2}\\
&=\frac{1}{2^N}\sum_{j=j_{\rm min}}^{N/2}\nu(j)(2j+1)g^{[N]}(j)\frac{\openone_{B_i}}{2}=\frac{\openone_{B_i}}{2}.
\end{align*}
Namely, although the optimal $|\psi\rangle$ is non-maximally entangled
in the $A_1A_2\cdots A_N:B_1B_2\cdots B_N$ cut, each $B_i$ qubit is still
maximally entangled with the other qubits, with both $A$ and $\bar B_i$ qubits,
in a complicated manner.
Note that this is also the case for the optimal $|\psi\rangle$ in the
deterministic version discussed in
Sec.\ \ref{subsec: Optimal state in the deterministic version}; we have
$\sigma_{B_i}=\openone/2$ by using Eq.\ (\ref{eq: deterministic X}), although
the optimal fidelity is nearly achieved by the maximally entangled
$|\psi\rangle$ as was shown in Fig.\ \ref{fig: fidelity}, in constrast to the
success probability.

%
\section{Example}
\label{sec: Example}

Now, let us show the explicit form of the optimal $|\psi\rangle$ and the
optimal POVM elements of Alice's measurement in the simplest case of $N=2$
in the probabilistic scheme. From Eq.\ (\ref{eq: Best probability}),
the optimal success probability in this case is $p=2/5$. From
Eq.\ (\ref{eq: probabilistic X}) for $N=2$, we have 
\begin{align*}
X&=\frac{2}{5}\openone(0)_{A}+\frac{6}{5}\openone(1)_{A}
\nonumber\\
&=\frac{2}{5}|\psi^-\rangle\langle\psi^-|
+\frac{6}{5}\Big(
|00\rangle\langle00|+|11\rangle\langle11|+|\psi^+\rangle\langle\psi^+|\Big),
\end{align*}
where $A=A_1A_2$ and $|\psi^\pm\rangle=(|01\rangle\pm|10\rangle)/\sqrt{2}$.
The optimal $|\psi\rangle$ is thus
\begin{align}
|\psi\rangle&=\sqrt{\frac{1}{10}}|\psi^-\rangle_{A}|\psi^-\rangle_{B_1B_2}
\nonumber\\
&+\sqrt{\frac{3}{10}}\Big(
|00\rangle|11\rangle
+|11\rangle|00\rangle
-|\psi^+\rangle|\psi^+\rangle\Big)_{AB_1B_2}.
\label{eq: optimal state}
\end{align}
From Eq.\ (\ref{eq: Best probabilistic measurement}) for $N=2$, we have
$\tilde\Theta_1=(4/5)\openone_{A_2}$, and hence
\begin{align}
\Pi_1&=
\sqrt{X^{-1}}(P^{-}_{A_1C}\otimes\tilde\Theta_{1A_2})\sqrt{X^{-1}}
\nonumber\\
&=\Big(|\eta^-\rangle\langle\eta^-|+|\eta^+\rangle\langle\eta^+|\Big)_{AC},
\label{eq: optimal Pi1}
\end{align}
where $\{|\eta^-\rangle,|\eta^+\rangle\}$ are orthogonal states given by
\begin{align*}
|\eta^-\rangle_{AC}=\sqrt{\frac{2}{3}}|x^-\rangle_{A}|0\rangle_C
+\sqrt{\frac{1}{3}}|00\rangle_{A}|1\rangle_C, \\
|\eta^+\rangle_{AC}=\sqrt{\frac{2}{3}}|x^+\rangle_{A}|1\rangle_C
-\sqrt{\frac{1}{3}}|11\rangle_{A}|0\rangle_C,
\end{align*}
with $|x^{\pm}\rangle=(1/2)(\pm|\psi^+\rangle+\sqrt{3}|\psi^-\rangle)$.
The POVM element $\Pi_2$ is given by $A_1\leftrightarrow A_2$ in
Eq.\ (\ref{eq: optimal Pi1}) [and thus, only $|x^{\pm}\rangle$ is replaced with
$(1/2)(\pm|\psi^+\rangle-\sqrt{3}|\psi^-\rangle)$].
It is then easily confirmed that
\begin{align*}
\langle\eta^-|\Big[|\psi\rangle\otimes \big(a|0\rangle+b|1\rangle\big)_C\Big]
&=\frac{1}{\sqrt{10}}\big(a|0\rangle+b|1\rangle\big)_{B_1}|1\rangle_{B_2},\\
\langle\eta^+|\Big[|\psi\rangle\otimes \big(a|0\rangle+b|1\rangle\big)_C\Big]
&=\frac{-1}{\sqrt{10}}\big(a|0\rangle+b|1\rangle\big)_{B_1}|0\rangle_{B_2},
\end{align*}
and hence
\begin{equation*}
\sqrt{\Pi_1} |\psi\rangle\otimes \big(a|0\rangle+b|1\rangle\big)_C
=\frac{1}{\sqrt{5}}|\psi_{\rm res}\rangle\otimes \big(a|0\rangle+b|1\rangle\big)_{B_1}.
\end{equation*}
Therefore, the state of the $C$ qubit is certainly teleported to the $B_1$
qubit faithfully, when Alice obtains $\Pi_1$ in her measurement 
(the coefficient on the right hand side represents the success probability of
$p/N=1/5$). Here,
\begin{align}
|\psi_{\rm res}\rangle&=\frac{1}{\sqrt{2}}\Big(|\eta^-\rangle_{AC}|1\rangle_{B_2}-|\eta^+\rangle_{AC}|0\rangle_{B_2}\Big) \nonumber\\
&=\frac{1}{\sqrt{2}}|\psi^-\rangle_{A}|\psi^-\rangle_{CB_2}\nonumber\\
&+\frac{1}{\sqrt{6}}\Big(
|00\rangle|11\rangle
+|11\rangle|00\rangle
-|\psi^+\rangle|\psi^+\rangle\Big)_{ACB_2}
\label{eq: residual state}
\end{align}
is the residual state after the teleportation is successfully completed.

%
\section{Entanglement consumption}
\label{sec: Entanglement consumption}

\begin{figure}[t]
\centerline{\scalebox{0.48}[0.48]{\includegraphics{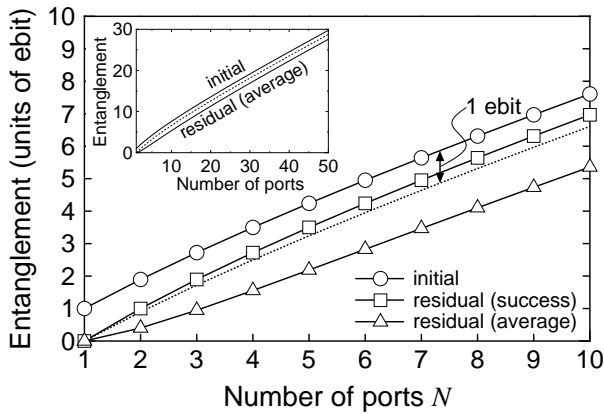}}}
\caption{The amount of entanglement as a function of number of output
ports ($N$) in the probabilistic scheme. The initial amount of the optimal
$|\psi\rangle$ (circles), the residual amount when the teleportation is
successfully finished (rectangles), and the average residual amount (triangles)
are plotted. The dotted line shows the eye-guide that corresponds to the
entanglement consumption by 1 ebit. The inset shows the same comparison of
the initial and average residual amount up to $N=50$.
}
\label{fig: consumption}
\end{figure}

Here, let us briefly discuss the entanglement properties in the probabilistic
scheme. In the explicit example for $N=2$ shown in the previous section,
Alice and Bob initially share the state $|\psi\rangle$ given by
Eq.\ (\ref{eq: optimal state}). Using
$\sigma_{A}=\hbox{tr}_{B_1B_2}|\psi\rangle\langle\psi|$, the amount of
the entanglement of $|\psi\rangle$ is calculated to be
$E_{\rm ini}=-\hbox{tr}\sigma_{A} \log_2(\sigma_{A})\approx 1.90$ ebits
(entanglement bits), which is less than the possible maximal amount of 2 ebits
for $N=2$ (the optimal $|\psi\rangle$ is non-maximally entangled
as mentioned in Sec.\ \ref{sec: Optimal success probability}).
When Alice obtains $\Pi_1$ in her measurement, the state of the $C$ qubit
is faithfully teleported to the $B_1$ qubit, i.e., the $B_1$ qubit is
used for receiving the teleported state. However, Bob still has the $B_2$
qubit, and as a result, Alice and Bob still share the residual state
$|\psi_{\rm res}\rangle_{ACB_2}$ given by Eq.\ (\ref{eq: residual state})
after the teleportation is completed. The entanglement of
$|\psi_{\rm res}\rangle$ (in the $AC:B_2$ cut) is calculated to be just
$E_{\rm res}=1$ ebit. Therefore, when the teleportation succeeds, only
$E_{\rm ini}-E_{\rm res}=0.90$ ebits are violated (or consumed), in spite that
a state of a single qubit is faithfully teleported.
Figure \ref{fig: consumption} shows such a comparison for general $N$, where
the entanglement of
$|\psi\rangle$ and $|\psi_{\rm res}\rangle$ is plotted by circles and
rectangles, respectively. It is found from the figure that the entanglement
consumption is less than 1 ebit even for $N>2$; rather, the amount of the
consumption gradually decreases for increasing $N$
($\approx 0.52$ ebits for $N=50$).

This implies that the entanglement between Alice and Bob even increases
if they try to teleport half of a maximally entangled state as in the
entanglement swapping (and if the teleportation is successfully finished).
This is because Alice and Bob newly share 1 ebit by the entanglement swapping,
while the entanglement consumption is less than 1 ebit as shown above. 
For this peculiar feature in our probabilistic teleportation scheme, the
use of the optimal $|\psi\rangle$ is crucial; if $|\psi\rangle$ is fixed to a
maximally entangled state as discussed in 
Sec.\ \ref{subsec: Maximally entangled state in the probabilistic version},
we have $E_{\rm ini}=N$
and $E_{\rm res}\le N-1$ because the number of Bob's qubits involved in
$|\psi_{\rm res}\rangle$ is $(N-1)$, and hence the entanglement consumption
in this case always satisfies $E_{\rm ini}-E_{\rm res}\ge1$
(in fact, $E_{\rm ini}-E_{\rm res}\approx1.009$ and 1.004 for $N=10$ and
$N=50$, respectively).
Note that the peculiar feature of the increase of entanglement, of course, does
not contradict the laws of entanglement, because the scheme discussed here is
the probabilistic one and the probabilistic increase of entanglement by local
operations and classical communication (LOCC) has not been prohibited by the
laws of entanglement.

Let us then evaluate the average amount of the residual entanglement. When
Alice obtains $\Pi_0$, the teleportation fails where the residual state
generally depends on the input state to be teleported. Moreover, if the input
state is a mixed state, the residual state is also a mixed state. Since the
evaluation of the entanglement for such a mixed state is a very hard task, let
us consider the worst case where the entanglement of the residual state 
when the teleportation fails is regarded to be zero. Using the success
probability $p$ of Eq.\ (\ref{eq: Best probability}), the average
residual entanglement in this worst case is given by $pE_{\rm res}$, which is
plotted by
triangles in Fig.\ \ref{fig: consumption}.
The corresponding average amount of the entanglement consumption
($E_{\rm ini}-pE_{\rm res}$) is roughly 2.2 ebits for $N=10$, and for
$N=50$ also (the inset of Fig.\ \ref{fig: consumption}).
In this way, in our probabilistic teleportation scheme, although Alice and
Bob must initially share much entanglement of ${\cal O}(N)$ ebits
(see the inset of Fig.\ \ref{fig: consumption}), only a few ebits are consumed
on average during the teleportation procedure. It may be said
that the most of the initial entanglement is only used as a working
space.

%
\section{Summary}
\label{sec: Summary}

In this paper, we have considered the scheme of quantum teleportation, where
Bob has multiple ($N$) output ports and obtains the teleported state by simply
selecting one of the $N$ ports. We investigated both deterministic version
and probabilistic version of the teleportation scheme aiming to teleport
an unknown state of a qubit, and analytically determined the optimal protocols.
All protocols shown in this paper can asymptotically achieve the perfect
teleportation
(i.e., faithful teleportation with unit success probability)
in the limit of $N\rightarrow\infty$.

In the deterministic version of the teleportation scheme, if the state
$|\psi\rangle$ employed for the
teleportation is fixed to a maximally entangled state, the optimal measurement
performed by Alice is the square-root measurement, where the optimal fidelity
is given by Eq.\ (\ref{eq: Optimal fidelity})
[or Eq.\ (\ref{eq: Asymptotic optimal fidelity})].
If both $|\psi\rangle$ and Alice's measurement are simultaneously optimized,
the generalized square-root measurement becomes optimal. The optimal fidelity
in this case is given by Eq.\ (\ref{eq: Best fidelity}).

In the probabilistic version, the optimal success probability is given by
Eq.\ (\ref{eq: Optimal probability}) [or 
Eq.\ (\ref{eq: Asymptotic optimal probability})] if $|\psi\rangle$ is fixed to
a maximally entangled state, and given by Eq.\ (\ref{eq: Best probability}) if
$|\psi\rangle$ is also optimized.
In contrast to the KLM scheme (and in contrast to
the deterministic version of our scheme also), the success probability
is considerably enhanced by optimizing $|\psi\rangle$; namely, the use of
the non-maximally entangled $|\psi\rangle$ provides a considerable benefit
than the use of the maximally entangled $|\psi\rangle$. Moreover, we showed
that the scheme is not inefficient concerning the entanglement resource,
because only a few ebits are consumed on average even for large $N$. If the
optimal  $|\psi\rangle$ is employed for the entanglement swapping, the amount
of entanglement even increases when the teleportation is successfully
completed.

Note finally that the form of the optimal fidelity
Eq.\ (\ref{eq: Best fidelity}) and the form of the optimal success probability 
Eq.\ (\ref{eq: Best probability}) are relatively simple (although the
corresponding optimal $|\psi\rangle$ and Alice's measurement are not).
Those are the achievable upper bounds in the general setting of selecting one
of $N$ qubits assisted by classical communication. In this paper, those bounds
were obtained by the direct optimization, but it will be important to study
further how those bounds of having the simple form are related to the
fundamental laws of physics. For instance, is it possible to derive those
bounds only from the no-signaling condition? This seems an intriguing and
important open problem.

%
\begin{acknowledgments}
This work was supported by the Special Coordination Funds for Promoting
Science and Technology.
\end{acknowledgments}
%
\appendix
%
\section{Proof of Eq.~(\ref{eq:eigenvalue_equation for N})}
\label{sec: Proof of eigenvalue_equation for N}

The proof is carried out by induction by noting that $\rho=\rho^{[N]}$ is
constructed recursively;
\begin{equation*}
\rho^{[N]}=\rho^{[N-1]}\otimes \frac{\openone_{A_{N}}}{2}
+\frac{\openone_{A_{1}}}{2} \otimes \dots \otimes \frac{\openone_{A_{N-1}}}{2}
\otimes P_{A_{N}B}^{-}.
\end{equation*}
The eigenvalue equation Eq.~(\ref{eq:eigenvalue_equation for N}) to be proved
is then rewritten as
\begin{equation} \label{eq:eigenvalue_equation}
\rho^{[N]}|\Psi^{[N]}(\lambda_{j}^{\mp};m) \rangle
=\lambda_{j}^{\mp}|\Psi^{[N]}(\lambda_{j}^{\mp};m)\rangle,
\end{equation}
where we attached a superscript $[N]$ to the eigenstates to emphasize the
relevant system size.
Moreover, we introduce the shorthand notation for the CG coefficients, 
\begin{equation*}
\langle j_{1},m_{1};j \rangle_{\pm}
=\textstyle\langle j_{1},m_{1},\frac{1}{2},\pm \frac{1}{2} | j,m_{1} \pm \frac{1}{2} \rangle,
\end{equation*}
and introduce $m_{\pm}=m\pm 1/2$, $m_{\pm \pm}=m\pm 1$ and similarly for $j$.

Since Eq.\ (\ref{eq:eigenvalue_equation}) is obvious for $N=1$, our aim is
reduced to proving Eq.~(\ref{eq:eigenvalue_equation})
under the assumption that Eq.~(\ref{eq:eigenvalue_equation}) with
$N\rightarrow N-1$ holds true. To this end, we write 
$| \Psi^{[N]} \rangle $ in terms of $| \Psi^{[N-1]} \rangle $ as follows.
\begin{widetext}
\begin{align} \label{eq:Psi_2I}
\lefteqn{|\Psi_{\mathrm{I}}^{[N]}(\lambda_{j}^{\mp};m)\rangle}
\quad\quad\nonumber \\
=&|\Psi^{[N-1]}(\lambda_{j_{+}}^{-};m_{+})\rangle |0\rangle_{A_{N}}
\big[
 \langle j_{+},m_{++};j_{++} \rangle_{-}^{*}
 \langle j,m_{+};j_{\pm} \rangle_{-}
 \langle j_{+},m_{++};j \rangle_{-}
+\langle j_{+},m;j_{++} \rangle_{+}^{*}
 \langle j,m_{-};j_{\pm} \rangle_{+}
 \langle j_{+},m;j \rangle_{-} \big]  \nonumber \\
+& |\Psi^{[N-1]}(\lambda_{j_{+}}^{-};m_{-}) \rangle
 |1\rangle_{A_{N}}
\big[
 \langle j_{+},m;j_{++} \rangle_{-}^{*}
 \langle j,m_{+};j_{\pm} \rangle_{-}
 \langle j_{+},m;j \rangle_{+}
+\langle j_{+},m_{--};j_{++} \rangle_{+}^{*}
 \langle j,m_{-};j_{\pm} \rangle_{+}
 \langle j_{+},m_{--};j \rangle_{+} \big] \nonumber \\
+& | \Psi^{[N-1]}(\lambda_{j_{+}}^{+};m_{+}) \rangle
 | 0 \rangle_{A_{N}}
\big [ 
 \langle j_{+},m_{++};j \rangle_{-}^{*}
 \langle j,m_{+};j_{\pm} \rangle_{-}
 \langle j_{+},m_{++};j \rangle_{-}
+\langle j_{+},m;j \rangle_{+}^{*}
 \langle j,m_{-};j_{\pm} \rangle_{+}
 \langle j_{+},m;j \rangle_{-} \big] \nonumber \\
+& | \Psi^{[N-1]}(\lambda_{j_{+}}^{+};m_{-}) \rangle
 | 1 \rangle_{A_{N}}
\big [
 \langle j_{+},m;j \rangle_{-}^{*}
 \langle j,m_{+};j_{\pm} \rangle_{-}
 \langle j_{+},m;j \rangle_{+}
+\langle j_{+},m_{--};j \rangle_{+}^{*}
 \langle j,m_{-};j_{\pm} \rangle_{+}
 \langle j_{+},m_{--};j \rangle_{+} \big],
\end{align}
and
\begin{align} 
\label{eq:Psi_2II}
\lefteqn{|\Psi_{\mathrm{II}}^{[N]}(\lambda_{j}^{\mp};m) \rangle}
\quad\quad  \nonumber \\
=& | \Psi^{[N-1]}(\lambda_{j_{-}}^{-};m_{+}) \rangle
 | 0 \rangle_{A_{N}} 
\big[
 \langle j_{-},m_{++};j       \rangle_{-}^{*}
 \langle j    ,m_{+} ;j_{\pm} \rangle_{-}
 \langle j_{-},m_{++};j       \rangle_{-}
+\langle j_{-},m     ;j       \rangle_{+}^{*}
 \langle j    ,m_{-} ;j_{\pm} \rangle_{+}
 \langle j_{-},m;j \rangle_{-} \big]  \nonumber \\
+& | \Psi^{[N-1]}(\lambda_{j_{-}}^{-};m_{-}) \rangle
 | 1\rangle_{A_{N}}
\big [
 \langle j_{-},m;j \rangle_{-}^{*}
 \langle j,m_{+};j_{\pm} \rangle_{-}
 \langle j_{-},m;j \rangle_{+}
+\langle j_{-},m_{--};j \rangle_{+}^{*}
 \langle j,m_{-};j_{\pm} \rangle_{+}
 \langle j_{-},m_{--};j \rangle_{+} \big ] \nonumber \\
+& | \Psi^{[N-1]}(\lambda_{j_{-}}^{+};m_{+}) \rangle
 | 0 \rangle_{A_{N}}
\big [ 
 \langle j_{-},m_{++};j_{--} \rangle_{-}^{*}
 \langle j,m_{+};j_{\pm} \rangle_{-}
 \langle j_{-},m_{++};j \rangle_{-}
+\langle j_{-},m;j_{--} \rangle_{+}^{*}
 \langle j,m_{-};j_{\pm} \rangle_{+}
 \langle j_{-},m;j \rangle_{-} \big] \nonumber \\
+& | \Psi^{[N-1]}(\lambda_{j_{-}}^{+};m_{-}) \rangle
 | 1 \rangle_{A_{N}}
\big [
 \langle j_{-},m;j_{--} \rangle_{-}^{*}
 \langle j,m_{+};j_{\pm} \rangle_{-}
 \langle j_{-},m;j \rangle_{+}
+\langle j_{-},m_{--};j_{--} \rangle_{+}^{*}
 \langle j,m_{-};j_{\pm} \rangle_{+}
 \langle j_{-},m_{--};j \rangle_{+} \big].
\end{align}
\end{widetext}
Equations~(\ref{eq:Psi_2I}) and (\ref{eq:Psi_2II}) are obtained by calculating the overlap between 
$| \Psi_{\mathrm{I}(\mathrm{II})}^{[N]} \rangle $ 
given by Eqs.~(\ref{eq:Psi_1I}) and (\ref{eq:Psi_1II}) and 
$| \Psi^{[N-1]} \rangle $ given by Eq.~(\ref{eq:eigenvector}) with $N \rightarrow N-1$.

The vector
$\rho^{[N-1]} \otimes \openone_{A_{N}}
|\Psi_{\mathrm{I}(\mathrm{II})}^{[N]}(\lambda_{j}^{\mp};m) \rangle $ 
takes the form of the right hand side of Eqs.~(\ref{eq:Psi_2I}) and (\ref{eq:Psi_2II})
with
\begin{align*}
&| \Psi^{[N-1]}(\lambda_{j_{+}}^{\mp};\dots) \rangle 
\rightarrow
\lambda_{j_{+}}^{[N-1]\mp}
| \Psi^{[N-1]}(\lambda_{j_{+}}^{[N-1]\mp};\dots) \rangle, \\
&| \Psi^{[N-1]}(\lambda_{j_{-}}^{\mp};\dots) \rangle 
\rightarrow
\lambda_{j_{-}}^{[N-1]\mp}
| \Psi^{[N-1]}(\lambda_{j_{-}}^{[N-1]\mp};\dots) \rangle,
\end{align*}
and these are further written in terms of $| \Phi^{[N-1]} \rangle $.
Here, we again attached a superscript $[N]$ to eigenvalues $\lambda^{\mp}_j$
to emphasize the relevant system size. On the other hand, the vector 
$\openone_{\bar A_{N}} \otimes P_{BA_{N}}^{-}
| \Psi_{\mathrm{I}(\mathrm{II})}^{[N]}(\lambda_{j}^{\mp};m) \rangle $ 
takes the form of the right hand side of Eqs.~(\ref{eq:Psi_1I}) and (\ref{eq:Psi_1II}) with
\begin{align*}
| 0 \rangle_{B}| 1 \rangle_{A_{N}} &\rightarrow
(
 | 0 \rangle_{B} | 1 \rangle_{A_{N}}
-| 1 \rangle_{B} | 0 \rangle_{A_{N}} ) /\sqrt{2}, \\
 | 1 \rangle_{B} | 0 \rangle_{A_{N}} &\rightarrow
-(
 | 0 \rangle_{B} | 1 \rangle_{A_{N}}
-| 1 \rangle_{B} | 0 \rangle_{A_{N}} ) /\sqrt{2}.
\end{align*}
Putting these two results together (and after lengthy calculations), 
we can see the desired eigenvalue equation, 
\begin{align*}
\lefteqn{\rho ^{[N]}| \Psi _{\mathrm{I}(\mathrm{II})}^{[N]}(\lambda_{j}^{\mp };m)\rangle} \quad\quad\\
=&\Big( 
\rho^{[N-1]}\otimes \frac{\openone_{A_{N}}}{2}
+\frac{\openone_{\bar A_{N}}}{2^{N-1}} \otimes P_{A_{N}B}^{-} \Big)
|\Psi _{\mathrm{I}(\mathrm{II})}^{[N]}(\lambda _{j}^{\mp };m)\rangle  \\
=&\lambda_{j}^{\mp}| \Psi_{\mathrm{I}(\mathrm{II})}^{[N]}(\lambda_{j}^{\mp};m) \rangle .
\end{align*}
This completes the proof.

%
\section{Proof of Eq.\ (\ref{eq: Bounded fidelity})}
\label{sec: Proof of bounded fidelity}

The proof is based on the technique used in the Holeve-Schumacher-Westmoreland
(HSW) theorem \cite{Holevo98a,Schmacher97a}. Let us denote
the eigenstates of $\sigma^{(i)}$ by $|k^{(i)}\rangle$, and hence
$\sigma^{(i)}=(1/d^{N-1})\sum_k |k^{(i)}\rangle\langle k^{(i)}|$.
The entanglement fidelity then satisfies
\begin{align*}
F&=\frac{1}{d^2}\hbox{tr}\sum_{i=1}^N \rho^{-1/2} \sigma^{(i)} \rho^{-1/2}
\sigma^{(i)} \\
&\ge\frac{1}{d^{2N}}\sum_{i=1}^N \sum_{k}|\langle k^{(i)}|\rho^{-1/2}|k^{(i)}\rangle|^2 \\
&=\frac{1}{Nd^{N-1}}\sum_{i=1}^N \sum_{k}|\langle k^{(i)}|\Big(d^{N+1}\rho/N\Big)^{-1/2}|k^{(i)}\rangle|^2 \\
&\ge\frac{2}{Nd^{N-1}}\sum_{i=1}^N \sum_{k}\langle k^{(i)}|\Big(d^{N+1}\rho/N\Big)^{-1/2}|k^{(i)}\rangle-1 \\
&=\frac{2}{d^{N+1}}\hbox{tr}\Big(d^{N+1}\rho/N\Big)^{1/2}-1
\ge 2-\frac{d^{N+1}}{N^2}\hbox{tr}\rho^2
\end{align*}
where $x^2\ge 2x-1$ was used in the second inequality, and 
$2(\openone-\Gamma^{1/2})\le 2\openone-3\Gamma+\Gamma^2$ was used
in the last inequality \cite{Holevo98a,Schmacher97a,NielsenChuang}. Since
\begin{align*}
\hbox{tr}\rho^2=\frac{1}{d^{N-1}}\sum_{i,j=1}^{N}
\langle k^{(i)}|\sigma^{(j)}|k^{(i)}\rangle
=\frac{N}{d^{N-1}}+\frac{N(N-1)}{d^{N+1}},
\end{align*}
we have $F\ge1-(d^2-1)/N$, and thus Eq.\ (\ref{eq: Bounded fidelity}),
because $f=(Fd+1)/(d+1)$ \cite{Horodecki99a}.

%
\section{Optimality of Eq.\ (\ref{eq: Best fidelity})}
\label{sec: Best fidelity}

The problem of maximizing $F$ given by Eq.\ (\ref{eq: F}) under the constraints
of Eqs.\ (\ref{eq: POVM}) and (\ref{eq: Normalization}) is a semidefinite
program \cite{Boyd04a} and thus has the dual problem. Since the Lagrange
function is 
\begin{align*}
{\cal L}&=
\sum_{i=1}^N \hbox{tr}\tilde \Pi_i \sigma^{(i)}
-\hbox{tr}\Omega\Big(\sum_{i=1}^N \tilde \Pi_i \!-\! X\otimes \openone\Big)
-a(\hbox{tr}X\!-\!2^N) \\
&=2^N a-\sum_{i=1}^N \hbox{tr}\tilde \Pi_i (\Omega\!-\!\sigma^{(i)})
-\hbox{tr}X(a\openone\!-\!\hbox{tr}_B \Omega),
\end{align*}
where $\Omega$ and $a$ are the Lagrange multipliers, the dual problem is of
minimizing $F=2^{N-2}a$ subject to
\begin{equation}
a\openone_A-\hbox{tr}_B \Omega\ge0, \hbox{~~} \Omega-\sigma^{(i)}\ge0.
\label{eq: Dual constraint of the best fidelity}
\end{equation}
Let us take 
\begin{align*}
a&=\frac{1}{2^{N-2}}\cos^2\frac{\pi}{N+2}, \\
\Omega&=\frac{1}{2^{N-1}}
\sum_{s=s_{\rm min}}^{(N-1)/2} c(s,y(s))\rho(s)^{1/y(s)}.
\end{align*}
Since any feasible solution of the dual problem gives an upper bound of the
original problem \cite{Boyd04a}, and $F=2^{N-2}a$ agrees with
Eq.\ (\ref{eq: Best entanglement fidelity}), it is then enough to show that
the above $\Omega$ is a feasible solution, i.e., $\Omega$ satisfies the
constraints of Eq.\ (\ref{eq: Dual constraint of the best fidelity}). It is
found from Eq.\ (\ref{eq:eigenvector}) and the CG coefficients that
\begin{align*}
\lefteqn{2^{N-1}\hbox{tr}_B\Omega=} \quad \nonumber \\
& c(s_{\rm min})
\big(\lambda_{s_{\rm min}-1/2}^-\big)^{1/y(s_{\rm min})}
\frac{2s_{\rm min}+1}{2s_{\rm min}} \openone(s_{\rm min}-{\textstyle\frac{1}{2}})_A \nonumber \\
+&\sum_{s=s_{\rm min}+1}^{(N-1)/2}
\Big\{c(s)\big(\lambda_{s-1/2}^-\big)^{1/y(s)}\frac{2s+1}{2s} \nonumber \\
&\quad+c(s-1)\big(\lambda_{s-1/2}^+\big)^{1/y(s-1)}\frac{2s-1}{2s}\Big\}
\openone(s-{\textstyle\frac{1}{2}})_A \nonumber \\
+&c({\textstyle\frac{N-1}{2}})
\big(\lambda_{N/2}^+\big)^{1/y({\textstyle\frac{N-1}{2}})}\frac{N}{N+1}
\openone({\textstyle \frac{N}{2}})_A,
\end{align*}
where $c(s)\equiv c(s,y(s))$ and $\openone(j)$ is the identity on the
subspace spanned by $|\Phi^{[N]}(j,\cdots)\rangle$. Since
\begin{align*}
\lefteqn{c(s)\big(\lambda_{s-1/2}^-\big)^{1/y(s)}\frac{2s+1}{2s}}
\quad\quad \\
\lefteqn{+c(s-1)\big(\lambda_{s-1/2}^+\big)^{1/y(s-1)}\frac{2s-1}{2s}} \\
&\quad =1+\frac{D(s)}{2}+\frac{1}{2D(s-1)}=2\cos^2\frac{2\pi}{N+2},
\end{align*}
we have $\hbox{tr}_B\Omega=a\openone_A$, and hence the first constraint in
Eq.\ (\ref{eq: Dual constraint of the best fidelity}) is satisfied.
Moreover, in the same way as in Ref.\ \cite{Ishizaka08b},
\begin{equation*}
\rho(s)^{1/y(s)}-\frac{1}{c(s,y(s))}\sum_{m,\beta}|\xi^{(i)}(s,m,\beta)\rangle\langle\xi^{(i)}(s,m,\beta)|\ge0
\end{equation*}
follows from Eq.\ (\ref{eq:matrix_element}), and thus the second constraint
in Eq.\ (\ref{eq: Dual constraint of the best fidelity}) is also satisfied.

%
\section{Optimality of Eq.\ (\ref{eq: Optimal probability})}
\label{sec: Optimal probability}

The problem of maximizing $p$ given by Eq.\ (\ref{eq: p2}) under the
constraints of Eqs.\ (\ref{eq: Theta}) (with fixed $X=\openone$) is also a
semidefinite program. The dual problem is of minimizing
$p=(1/2^{N+1})\hbox{tr}\Omega$ subject to
\begin{equation}
\Omega\ge0, \hbox{~~}
\hbox{tr}_{A_iB} P^-_{A_iB} \Omega_{AB} \ge \openone_{\bar A_i}.
\label{eq: Dual constraint of the optimal probability}
\end{equation}
For the choice of
\begin{equation*}
\Omega=\sum_{s=s_{\rm min}}^{(N-1)/2}
\frac{2s+1}{s+1}\sum_{m,\beta}
|\Psi(\lambda_{s+1/2}^+;m)\rangle\langle\Psi(\lambda_{s+1/2}^+;m)|,
\end{equation*}
it is found from Eqs.\ (\ref{eq: overlap 3}) and (\ref{eq: overlap 4}) that
\begin{equation*}
\hbox{tr}_{A_iB}P^-_{BA_i}\Omega=\sum_{s=s_{\rm min}}^{(N-1)/2}
\frac{2s+1}{s+1}\frac{s+1}{2s+1}\openone(s)_{\bar A_i}=\openone_{\bar A_i},
\end{equation*}
and therefore the above $\Omega$ is a feasible solution.
On the other hand, $p=(1/2^{N+1})\hbox{tr}\Omega$ agrees with 
Eq.\ (\ref{eq: Optimal probability}), because the degeneracy of
$|\Psi(\lambda_{s+1/2}^+;m)\rangle$ is $(2s+1)g^{[N]}(s+1/2)$
[see Eq.\ (\ref{eq: degeneracy +})] and
\begin{equation*}
\frac{(2s+1)^2 g^{[N]}(s+1/2)}{s+1}=
\frac{N}{2^{N-1}}\frac{(2s+1)g^{[N-1]}(s)}{\lambda_{s+1/2}^+}.
\end{equation*}

%
\section{Optimality of Eq.\ (\ref{eq: Best probability})}
\label{sec: Best probability}

The dual problem for general $X$ is of minimizing
$p=2^Na$ subject to
\begin{equation}
\Omega\ge0, \hbox{~~}
\hbox{tr}_{A_iB} P^-_{A_iB} \Omega_{AB} \ge \openone_{\bar A_i},
\hbox{~~} a\openone_A-\frac{1}{2^{N+1}}\hbox{tr}_B\Omega\ge0.
\label{eq: Dual constraint of the best probability}
\end{equation}
Let us take $a=(1/2^N)N/(N+3)$ and consider
\begin{align*}
\Omega=\sum_{s=s_{\rm min}}^{(N-1)/2}\sum_{m,\beta}
\Big\{
&d(s)|\Psi(\lambda_{s+1/2}^+;m)\rangle\langle\Psi(\lambda_{s+1/2}^+;m)| \\
+&e(s)|\Psi(\lambda_{s-1/2}^-;m)\rangle\langle\Psi(\lambda_{s-1/2}^-;m)|\Big\},
\end{align*}
where $d(s)=(N+3+2s)/(N+3)$ and $e(s)=(N+1-2s)/(N+3)$.
Since $p=2^Na$ agrees with Eq.\ (\ref{eq: Best probability}), it is enough to
show that the above $\Omega$ is a feasible solution. From
Eqs.\ (\ref{eq: overlap 1}), (\ref{eq: overlap 2}), (\ref{eq: overlap 3}),
and (\ref{eq: overlap 4}), it is found that
\begin{align*}
\hbox{tr}_{A_iB}P^-_{A_iB}\Omega&=\sum_{s=s_{\rm min}}^{(N-1)/2}
\Big\{\frac{d(s)(s+1)}{2s+1}
+\frac{e(s)s}{2s+1}\Big\}\openone(s)_{\bar A_i} \\
&=\openone_{\bar A_i},
\end{align*}
and the second constraint of
Eq.\ (\ref{eq: Dual constraint of the best probability}) is satisfied.
Moreover, it is found from Eq.\ (\ref{eq:eigenvector}) and the CG
coefficients that
\begin{align*}
\hbox{tr}_B\Omega=\sum_{s=s_{\rm min}}^{(N-1)/2}\Big\{ 
&\frac{d(s)(2s+1)}{2(s+1)}\openone({\textstyle s+\frac{1}{2}})_A \\
+&\frac{e(s)(2s+1)}{2s}\openone({\textstyle s-\frac{1}{2}})_A \Big\}
=\frac{2N}{N+3}\openone_A,
\end{align*}
and hence the third constraint of
Eq.\ (\ref{eq: Dual constraint of the best probability}) is also satisfied.


%

\end{document}